# Global Population and Carrying Capacity in the Anthropocene: the Relative Growth Rate Insight

[1]Aleksandra Drozd-Rzoska, [2]Agata Angelika Sojecka, [1]Sylwester J. Rzoska

[1]Institute of High Pressure Physics of the Polish Academy of Sciences,
ul. Sokołowska 29/37, 01-142 Warsaw, Poland

[2]University of Economics in Katowice, Faculty of Management, Department of Marketing,
ul. 1 Maja 50, 40-257 Katowice, Poland

(*) Aleksandra Drozd-Rzoska is the corresponding author; e-mail: arzoska@unipress.waw.pl

**Abstract**

This report provides insights into global population dynamics since the beginning of the Anthropocene, focusing on empirical data and minimizing a priori the impact of model assumptions. It explores the Relative Growth Rate concept, introduced recently to global population studies by Lehman et al. (*PNAS* **118**, e2024150118 (2021)) and subsequently extended to its analytical counterpart (*PLoS ONE* **20**, e0323165 (2025)).

The analysis reveals a general non-monotonic growth pattern in the Anthropocene, emphasising the uniqueness of the Industrial Revolution era. For the first 290 years, the 'Doomsday' *critical* scaling provides an excellent description for population changes, with a singularity at 2026. This is followed by the crossover to an exceptional *'reversed criticality'* pattern, which has held over the last six decades, to the present day.
The analysis suggests that the evolution of the human Real Intelligence (RI) population during the innovation-driven Industrial Revolution — a period of rapidly increasing connectivity and complexity — can serve as a model counterpart for the puzzling dynamics of Artificial Intelligence (AI) growth.

The final conclusion is positive: the catastrophic Doomsday singularity can be avoided due to generic system constraints, both for RI and AI.

## 1. Introduction

In recent decades, global climatic and environmental threats, exacerbated by political instabilities, have posed catastrophic risks to modern civilisation [1-4]. More recently, the challenge of Artificial Intelligence (AI) has emerged. While its development is seen as an opportunity to enter a new civilizational stage, AI is also considered the most significant threat ever [4-6]. It has already 'reformed' - or maybe 'deformed' - many areas of life [6,7]. Its seemingly spontaneous growth has also significantly increased demands for energy and critical raw materials, thereby exacerbating the climate and environmental crises [5,6]. Perhaps no other topic in the history of Science has generated as much interest as AI. In 2024 alone, the number of reports reached an astonishing 420,000 [8].

However, the fundamental question of a general pattern of AI growth remains puzzling. This includes the essential question of how to reliably predict the future rise of AI [3-7].
The first question should concern the metric that reflects AI development. An option could be the number of information bits 'produced' on Earth annually. In 2020, Vopson [9] suggested that the power required to sustain this digital production will exceed $18.5 \times 10^{15} W$ in approximately 300 years. This inevitably leads to a singularity and the destruction of humanity, as it was highlighted in the title of his seminal work, '*The Information Catastrophe*' [9]. The generation and exchange of information, particularly in relation to AI and associated technologies such as blockchain finance, already consume vast amounts of energy and require ever-increasing quantities of raw materials. Considering the number of planned AI *'Giga-Factories'* [10] this may only be the beginning. Consequently, the AI boost could lead to a hypothetical global extinction scenario, especially when supported by feedback interactions with other mentioned catastrophic singularities [3-7].

When searching for Artificial Intelligence (AI) growth scaling patterns, one can consider the only explicitly known and highly successful case – human Real Intelligence (RI), particularly during the innovation-driven Industrial Revolution epoch. For its developmental metric, the fundamentally validated global population growth scaling pattern can be considered. The very recent explicit link between human population growth and AI changes in China, shown over the 2008-2020 period, [11] is worth recalling here.

This report presents the analysis of global population growth since the onset of the Anthropocene (12,000 BCE), with particular focus on the ongoing Industrial Revolution epoch. It is based on a unique analytical dataset of 206 population data, tested via the per capita Relative Growth Rate (RGR), that allowed for subtle, distortions-sensitive, insight with minimized external/heuristic model assumptions.

The RGR concept for global population studies was introduced only recently by Lehman et al. (2021, [12]) and subsequently developed into an analytic counterpart (2025, [13]).

So far, the global population growth has been analysed using two other cognitive paths [13]:

- The first approach involves defining the global population by examining the impact of geographical regions, social groups, education, and various social interactions. Particular focus is given to the roles of women and migration, as well as the birth/death ratio, age structure, and economic development. A multitude of data is analysed statistically using models developed in management, econometrics, bio-evolution, and socio-economics [14-18]. However, this dominant approach has two significant limitations. Firstly, precise data only became available after 1950. Secondly, the 'impact weights' of different factors often require subjective expert opinion. This may explain why population forecasts for 2100 range from 4 to 15 billion [19]. Notwithstanding, this approach is significant for tracking specific issues that affect population growth, such as fertility, mortality, aging, education, disease evolution, migration, and economic issues.

- The second approach uses a scaling model to describe changes in $P(t)$ directly. This is usually achieved using non-linear fitting routines [20-23]. However, significant challenges remain. Firstly, when shifting to the past, there is considerable scatter in the population data. Secondly, the fundamental validity of the scaling equation used is called into question, as is the arbitrariness of the selected testing period. Thirdly, validating the fair visual portrayal of 'empirical' data becomes problematic as the number of adjustable parameters increases, leading to enormous increases in their individual errors.

Nowadays, it is clear that despite decades of studies and their significance for global management and governance, the problem remains a challenging puzzle. It can be illustrated by the scatter of global population forecasts for 2100, ranging from 5 billion to 15 billion [19].

This report presents results of global population $P(t)$ studies via the model-limited approach, focused on distortions sensitive RGR-based analysis able to reveal subtle local temporal features. To deliberate on human Real Intelligence population growth as a possible pattern for AI general growth, the singular 'critical' scenario seems significant, particularly when recalling heuristic catastrophic expectations for AI development [3-7, 24-28].

It recalls the famous Doomsday equation introduced by von Forster et al. (1960, [29]) for global population studies, with the exceptional catastrophic singularity:

$$P(t) = \frac{C}{(D-t)^M} \quad \Rightarrow \quad log_{10}P(t) + log_{10}C - Mlog_{10}(D-T) \tag{1}$$

where $P(t)$ is for temporal population changes; 'empirical' data analysis yielded $C = 179\ billion$, $D = 2026.87$ is for the 'Doomsday' singularity with infinite population, and the exponent $M = 0.99$; related errors were not explicitly stated; the right part of Eq. (1) is for the log-log scale analysis with adjustable $D$ parameters used in ref. [29].

The authors of ref. [29] showed the excellent representation of $P(t)$ data for the enormous period from 400 CE to 1958. Von Forster et al. [29] used the 24 'historical' population data available at the time. Equation (1) is commonly named the '*Doomsday Equation*' or the '*Hyperbolic Equation*, since one can assume $M = 1$ for the power exponent [20-23].

Notably, this impressive result inspired researchers across a range of fields, including biology, ecology, sociology, economics, and political science [30-35]. However, the discovery was also quickly and essentially criticised by demographic researchers [36-46].

The criticism aimed at the 'exotic' prediction of human extinction in 2026, the lack of a model support, and the 'suspicious simplicity'. The criticism increased when more global population data became available, and a scatter of exponent values $0.7 < M \leq 1$ was reported [42-44]. Finally, von Foerster et al.'s [29] scaling relation was 'relegated to the periphery' in global demographic studies when the rising divergence from real-world population data became evident since the late 1970s [20-23, 31,32,45-47].

This report shows that, despite these objections above, the 'critical' Doomsday scenario offers the optimal portrayal of $P(t)$ changes during the first 290 years of the Industrial Revolution, as validated by RGR-based analysis and presented below. However, the catastrophic singularity is avoided due to the crossover ('tunnelling') to the unique *reversed-criticality* pattern that has occurred over the last 6 decades to date.

This result can be optimistic for the human Real Intelligence (RI) global population and its Artificial Intelligence (AI) counterpart: the 'Doomsday' can be (is?) avoided.

## 2. Relative Growth Rate analysis of population dynamics

In 2021, Lehman et al. [12] presented a challenging discussion of Verhulst-type multiparameter global population modeling since the onset of the Anthropocene, with relevant parameters linked to overcoming subsequently emerging ecological barriers. The essential significance of the per capita Relative Growth Rate (RGR, $G_P$) for the basic validation of model analysis was presented. RGR was introduced in the discrete form, enabling the iterative analysis of population comparing population steps-changes $\Delta P(t_i)$ in subsequent time periods $\Delta t_i$, namely [12,13]:

$$P(t) \Rightarrow G_P(t_i, P_i) = \frac{1}{P_i}\frac{\Delta P(t_i)}{\Delta t_i} \Rightarrow G_P(t) = \frac{dP(t)/dP}{dt} = \frac{d\ln P(t)}{dt} \qquad (2)$$

where $P_i = P(t_i)$ is the reference population for subsequent times steps $t_i$.

The left side of this relation is for the original definitions given in ref. [12], and the right-hand side is for the RGR analytical counterpart developed in refs. [13] to test super-Malthusian and extended Verhulst global population scaling.

Returning to the 'catastrophic' von Foerster et al. (1960, [29]) Eq. (1) description of the global population growth, the criticism appeared almost immediately after announcing the results. It took place despite the unique features: (*i*) the simple functional form and easy application, (*ii*) the unique high-quality reproduction of empirical data over the extreme period ~1,500 years. The criticism focused on the 'strange' forecast of human extinction when approaching 2026, associated with the earlier global exhaustion of resources. The lack of model justification was often stressed, although ref. [29] presented only a first communication [36-46]. These objections, supported by mounting evidence of discrepancies compared with real population data, meant that von Foerster et al. [29] ceased to consider Eq. (1) relevant (or reliable) for describing global population changes.

This work validates the opposite conclusion: the 'Doomsday equation' provides the preferable scaling of the global population changes over the majority of the Industrial Revolution epoch, approximately over the period of 290 years.

It is worth noting that the 'model justification' of the Doomsday Eq. (1) could be found already in the mid-1970s, via the following differential equation proposed by Caswell [48]:

$$\frac{dP(t)}{dt} = r(t)[P(t)]^{1+1/m} \Rightarrow G_P(P,t) = \frac{1}{P(t)}\frac{dP(t)}{dt} = \frac{d\ln P(t)}{dt} = r(t)[P(t)]^{1/m} \qquad (3)$$

The left part, with $r(t) = r = const$, is for the original Caswell dependence [48]. The right part presents the authors' of the given report formulation using the analytic RGR factor.

For the parameter $m = 1$, the solution of Eq. (3) leads to the '*Hyperbolic, Doomsday*' Eq. (1). For $m \to \infty$ one obtains the differential equation coupled to the basic Malthus (1798, [49]), where the population rise is proportional to the current population itself. It is related to $G_P = const$ and $r(t) = r = const$, namely:

$$\frac{dP(t)}{dt} = rP(t) \Rightarrow \frac{dP/P}{dt} = \frac{d\ln P(t)}{dt} = G_P(t) = r \qquad (4)$$

where the left side of Eq. (4) is for the original to the Malthus model [49], and its right-hand parts shows the RGR-based form introduced by the authors – the extended Caswell relation. The above relation leads to the following 'classic' Malthusian exponential equation describing the population growth [49]:

$$P(t) = P_0 exp(rt) \rightarrow log_{10}P(t) = log_{10}P_0 + rt/ln10 \qquad (5)$$

where the right-hand part shows the log-log scale linearization.

Notably, the exponential rise defined by Eq. (5) predicts essentially a 'weaker' rise than the hyperbolic growth defined by Eq. (1).

One can further consider the following transformation of the right-hand part of Eq. (3):

$$G_P(P) = r[P(t)]^{1/m} \quad \Rightarrow \quad log_{10}G_P = log_{10}r + \frac{1}{m}log_{10}P(t) \qquad (6)$$

For the plot $log_{10}G_P$ vs. $log_{10}P(t)$, the sloped linear domain validate description via Eq. (1) manifest by a linear dependence. If $K = M = 1$, it is related to the 'hyperbolic' behaviour. For $K \neq 1$, one should expect the exponent $M \neq 1$. The horizontal line behaviour ($G_P(P) = const$) indicates the underlying simple Malthusian pattern.

For the direct distortions – sensitive test of the time-dependent Eq. (1), with the arbitrary exponent $M$, one can propose the following analysis:

$$P(t) = \frac{C}{(D-t)^M} \Rightarrow lnP(t) = lnC - Mln(D-t) \Rightarrow G_P(t) = \frac{dlnP(t))}{dt} = \frac{-M}{D-t} \Rightarrow$$

$$\Rightarrow G_P^{-1}(t) = \left[\frac{dlnP(t))}{dt}\right]^{-1} = -M^{-1}D - M^{-1}t = d - mt \qquad (7)$$

where $d = -M^{-1}D$, $m = M^{-1}$, and the singularity is related to the condition $G_P^{-1}(t = D) = 0$. The plot $G_P^{-1}(t)$ vs. $t$ validates the description of population changes in the time domain, also yielding optimal values of relevant parameters, with reliable errors estimations.

Concluding, one can transform $P(t)$ 'empirical' to $G_P(P,t)$ form to reveal and tests the dominant pattern of changes, to reveal preferences for the portrayal via a given model equation, without any preliminary assumptions. For the Doomsday Eq. (1) transformed following Eq. (7), it should follow a linear dependence. The discussion related to the Super-Malthus [49] and extended-Verhulst [50] scaling is presented in refs. [13,47,51]. Notably, optimal parameter values are also obtained, and nonlinear fitting is not required.

It should be emphasised that using Eq. (1) to describe the phenomenon associated with available empirical data significantly distant from the singularity — as in the case of the Doomsday $D$ in Eq. (1) — has to inevitably result in significant errors in the fitted parameters, despite the model's apparent ability to depict the data accurately. This problem is well evidenced in studies of pseudospinodal behaviour in near-critical liquids [52] and the previtreous domain of glass-forming systems [53]. Hence, applying an approach that parallels the analysis via Eq. (7) can be significant for these physics and material engineering issues. Notably, the parallel of Eq. (7) has already been successfully implemented to reveal critical-like patterns of previtreous changes in configurational entropy and heat capacity for glass-forming systems [53].

When discussing the dynamics described by Eq. (1), it is interesting to note the similarity to patterns observed in *Critical Phenomena Physics* [54]. This was a grand, significant achievement in 20th-century physics, as it explained the universal behaviour observed near a continuous or semi-continuous phase transition. For any material system in the near-critical (upper/sub-critical) domain, physical properties depend on the distance from the critical singular point and follow a power-law pattern. This behaviour is governed by a power exponent that depends solely on the general properties of the system: the space ($d$) and the order parameter ($n$) dimensionalities. This unusual characteristic has led to the (super/sub) critical region being considered a new state of matter characterised by the spontaneous formation of multimolecular/multispecies assemblies whose size and lifetime increase singularly towards infinity, as a critical point is approached [54]. Although Critical Phenomena Physics initially concerned only material systems, similar patterns were found in qualitatively different areas, ranging from quantum systems [55] to cosmology [56] and socio-economic issues [57]. The latter has given rise to a new field of Science: *Econophysics*.

The critical pattern of changes corresponds precisely to the Doomsday Eq. (1), suggesting that the singularity can represent a critical point, i.e., $D = T_C$. Spontaneous self-assembly is an inherent human tendency, and then it can be considered in the frames of Human Intelligent Soft Matter [47]. Initially, the correlation length of the spontaneously formed critical assemblies is small. For humans, these could be families and tribal groups. However, civilisation development involves increasingly complex interactions and significant cultural, socio-economic, and political links that create rising super-local links, and further coherent arrangements and orderings, i.e., the rise of the correlation length. Today, the correlation length approaches the spatial constraint of the size of our planet, Earth. Maybe these spatial constraints will be removed as the expansion encompasses the Solar System. But it is a melody of the future.

In 1920, Pearl and Reed [58] introduced the concept of carrying capacity ($K$) as a metric of resources available for the given population when discussing an extension of the Verhulst model [50] equation. In recent decades, this concept has also been widely discussed in the context of the emerging sustainable society as a response to the aforementioned catastrophic threats, where it is rather seen as a metric of emerging constraints/threats [12 and refs. therein]. In the opinion of the authors, the Carrying Capacity represents the interplay of the two mentioned factors. In such frames, worth recalling is the Condorcet equation introduced by Cohen (1995, [59]), which indicates that Carrying Capacity ($K$) changes can be coupled to Global Population changes:

$$\frac{dK(t)}{dt} = c(t)\frac{dP(t)}{dt} = \left(\frac{L}{P(t)}\right)\frac{dP(t)}{dt} \Rightarrow \tag{8}$$

$$\Rightarrow L\frac{dP(t)/P(t)}{dt} = L\frac{d\ln P(t)}{dt} = L \times G_P(t) \tag{9}$$

where $L=const.$

Eq. (8) is for the original Cohen's dependence [59], and Eq. (9) is for the RGR link introduced by the authors in ref. [47]. Cohen [59] indicated that the parameter $c$ in Eq. (8) can be decisive for the functional form describing carrying capacity changes, namely: (*i*) $c = 1$ was for the Malthus [49] exponential function (Eq. 5), (*ii*) for $c < 1$ the Verhulst behaviour [50] with the final stationary stage can emerge, and (*iii*) for $c > 1$ the catastrophic von Foerster et al. [29] pattern (Eq. (1)) can appear. Using equations (8) and (9), one can consider parameter $L$ L to represent the maximum possible population in the considered systems, given the available resources for the given stage of development, and the impact of emerging constraints.

### 3. RGR-based test of 'Doomsday', critical, global population modeling

Figure 1 presents 206 global population data spanning 12 millennia of the Anthropocene, the formative period of human civilization [60]. It is based on an analytical data set-up derived from numerical filtering of 'naturally' scattered data from different sources, as described in the Methods section. Notable are local characteristic changes that can be correlated with some civilizational and historical epochs and circumstances. For the applied semi-log scale, the basic Malthusian behaviour, related to geometric population growth, follows a linear pattern (Eq. (5)). When comparing the differential equations underlying Verhulst [12,50,58,59] and Malthus [49,20-22] models, the latter can be related to the 'infinite resources', but adjusted for rising population. In the Anthropocene, over the first 9 millennia, a simple Malthusian pattern is visible in Figure 1. This extreme period can be divided into two apparent Malthusian domains. After ~5,500 BCE, the growth rate increased ~4.5 times, as evidenced by the slope change in Figure 1. The crossover between these domains coincides with the transition from the Early to the Late Neolithic period [60]. For the authors, the significant rise in growth rate occurred relatively soon after the submersion of the Doggerland region, which connected present-day Great Britain with the rest of Europe and Scandinavia. The North Sea with an average depth of ~87 *meters* appeared [61]. This suggests a rapid influx of water into the world's oceans, which is possible only if the grand ice sheets melt and retreat quickly to the far North. It likely influenced climate change and could have increased food resources for Neolithic settlements. Subsequently, the simple Malthusian pattern, but with a significantly larger growth rate $r$, appears between ~1200 BCE and 800 BCE, i.e., from the late Bronze Age to the early Antiquity [60].

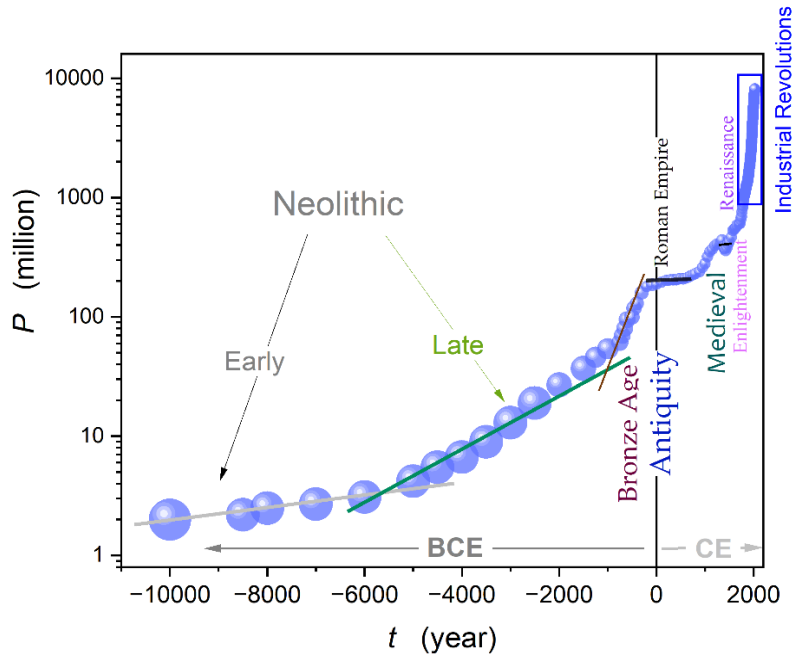

**Fig. 1 The global population change from the Anthropocene onset to 2024.** The semi-log scale visualizes simple geometric (Malthusian, Eq. 5) growth as straight, sloped lines. Successive historical epochs are marked. Horizontal lines indicate periods when the global population was approximately stable. They are related to the Roman Empire (Antiquity) and the Black Death pandemic, since the Late Medieval, times. The size of the point reflects the uncertainty. The plot is based on the analytical set developed by the authors as described in the Method section and given in the Supplementary Information.

Then the transition to a nearly constant global population period ~200 $million,$ lasting over 500 years, is visible. At that time, almost a quarter of the global population lived in the Roman Empire [62]. Possible reasons for this unique pattern are discussed in refs. [13,47,51] It is worth stressing that a similar impact on the global population had the Black Death pandemic [63] in the late Medieval and early Renaissance times, as visible in Fig. 1 and even stronger in Fig. 2. Since the fall of the Western Roman Empire (~500$CE$) [62], the systematic global population growth began. It follows the nonlinear, increasingly 'steeper', visible in the log-log scale presentation of Figure 1. In refs. [13,47] it was named the 'Super-Malthus' behavior, and discussed via super-Malthus (SM) equations, using the relaxation time $\tau = 1/r$.

Figure 2 presents the analysis of $P(t)$ data shown in Fig. 1 via the distortions-sensitive RGR $G_P(P)$ factor, defined via Eq. 2. The applied log-log scale facilities insight when multi-decade changes in tested values takes place. It also links the analysis to Eq. (6), which enables the test of the extended Caswell relation (Eq.(3)).

The distortions-sensitive analysis in Figure 2 explicitly shows that global population changes are essentially non-monotonic. Notably are two dominant scaling patterns:

- ✓ The Malthusian behavior (Eq. (4) and (5)) manifested via horizontal lines that appear for the Late Neolithic ($LN$, 5500 – 1700 BCE), Early Antiquity ($EA$, 700 – 200 BCE) and from Late Medieval to Baroque end ($MBq$, 1100 - 1680) periods.
- ✓ The 'Doomsday', critical–type rise manifested via positively sloped lines appears for the Early Neolithic (EN: 10 000 – 5 500 BCE), Bronze and Early Iron Ages (BrI: 1500 – 700 BCE), and in the Industrial Revolution era from 1680 to 1966.

For the Malthus behaviour RGR factor is equal to the Malthus growth rate, i.e., $G_p = r$: (Eq. 4). This enables the comparison of the Malthusian rise factor in the above epochs: $r(LN)/r(MBq) \approx 0.3$ and $r(EA)/r(MBq) \approx 1.2$.

Notably, Figure 2 reveals a non-Malthusian pattern for the Early Neolithic, which is hidden in the basic population data presentation in Figure 1.

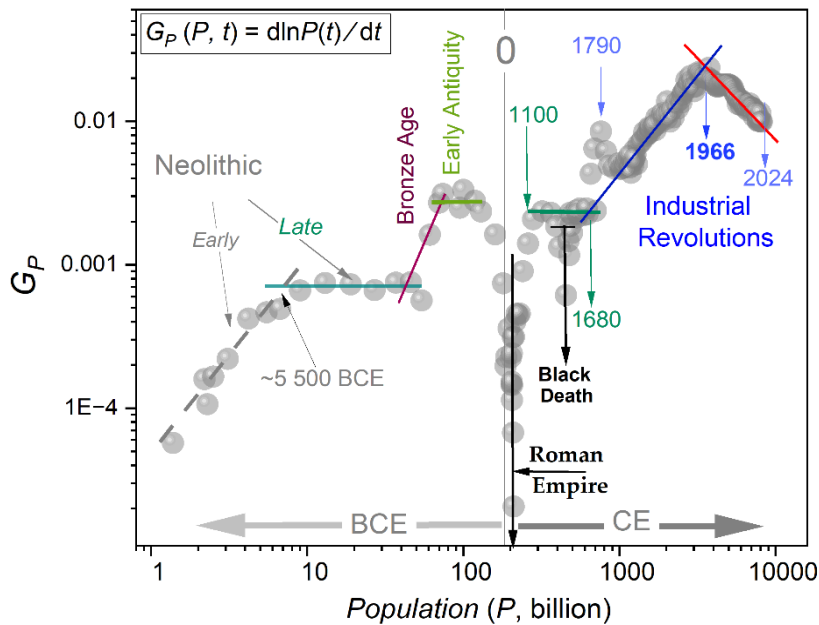

**Fig. 2 The log-log scale presentation of the relative growth rate (RGR) for the global population in Anthropocene**, based on population data shown in Fig. 1 and calculated as defined in Eq. (2). Emerging manifestations of historical 'events' and epochs are indicated. Note the link to Eq. (6).

The non-Malthusian, critical-like rise in global population during the Early Neolithic period can be attributed to the enormous resources, favourable climatic changes, and the Carrying Capacity of that epoch. The human population could utilise available resources relatively easily, which stimulated significant population growth. The emergence, implementation, and rapid dissemination of innovations were certainly crucial for raising the system's Carrying Capacity associated with desirable resources. According to the authors, pre-technologies of food preservation could be particularly important, as they enabled independence from ongoing hunting and gathering activities and from seasonal climatic changes. This could be the stacking of collected plants, a practice that might have originated from observations of animal behaviour. Furthermore, meats could be dried, fumigated, and perhaps even smoked. Notably, traces of pre-yogurt and pre-cheese appear during this period, indicating the progressive domestication of some animals and the ability to process milk for long-term use [60,64]. The accumulation of these factors, along with the tendency towards self-assembly, which is likely characteristic of both complex systems and humans with a uniquely broad and multi-dimensional range of interactions, led to the emergence of communities that engaged in agriculture [60] as a more stable food source. However, each new member of the population now has to acquire land independently to feed themselves and their families. This naturally leads to an increase in resources, in a feedback loop with population growth, and can eventually subsequently lead to a more reduced 'geometric' Malthusian rise in population.

It is explicitly visible in Figure 2 that after the years 1680 -1690, a qualitatively new pattern for the global population dynamics emerged. This is the Industrial Revolution era [47,65,66]. For the first period up to 1790, disturbances still occur, but later, up to ~1966, a nearly linear pattern with the exponent $K \approx 1$ for $log_{10}G_P(P)$ vs. $log_{10}t$ plot presented in Figure 2 appears. According to Eqs. (3 & 6) it directly supports the '*Doomsday, Critical*' von Foerster et al. [29] behaviour (Eq. 1), in the simplest hyperbolic form.

This is confirmed in Figure 3, via the explicit linear behaviour of RGR factor reciprocal $G_P^{-1}(t)$; shown in the inset, in agreement with Eq. (7).

Figures 2 and 3 present the validated evidence for the unusual crossover in the global population changes, from 1680 to 2024:

$$\sim 1680 \; \rightarrow \; (D_1 - t)^{M=-1} \; \Rightarrow \; \sim 1966 \; \Rightarrow \; [\,(t - D_2)^{M=-1}]^{-1} = \; (t - D_2)^{M=1.1} \qquad (10)$$

where $D_1 \approx 2026$, and $D_2 \approx 1919$.

Thus, for almost 290 years, the basic hyperbolic form of von Foerster et al. [29] Eq. (1), has represented the dominant trend in global population growth, with the 'critical' singularity at $D_1 \approx 2026$. After ~1966, the crossover of the highly complex global population system to the 'Reversed Criticality' pattern occurs. It can be considered as the 'tail' of a singularity hidden in the past, at $D_2 \approx 1919$.

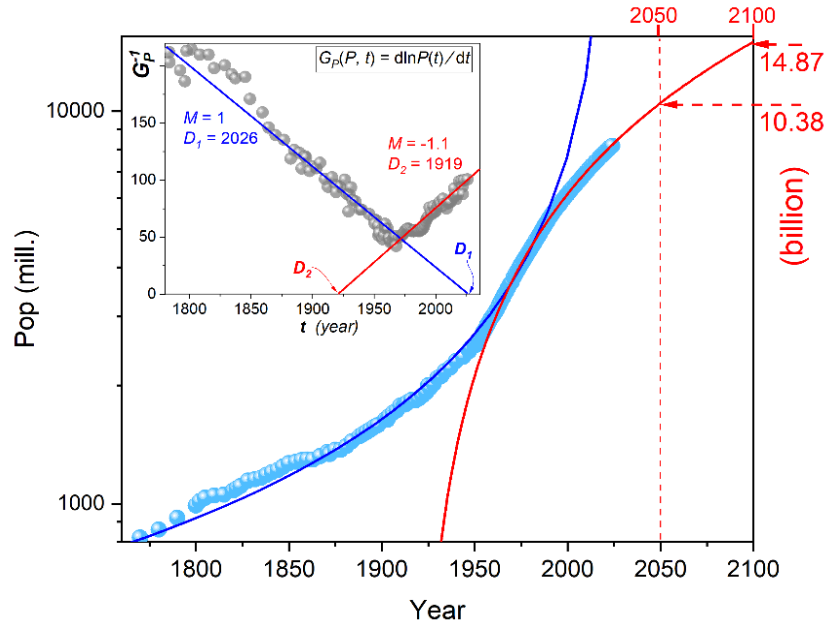

**Fig. 3 The focused insight into the global population growth in the Industrial Revolution era, since 1770**. The parameterizations via the Critical (related to $D_1 = 2026$ singularity) and the Reversed Critical (related to the singularity 'hidden' at $D_2 = 1919$) dependencies are shown, with respect to the definition given in Eq. (10). The validating RGR-related plot is shown in the insight: note the link to Eq. (7). It yields basic scaling parameters for the parameterization shown in the central part of the figure, via the relations included in Eq. (10).

It is related to the singularity hidden in the past, at $D_2 \approx 1919$, appears. After the crossover year, the global population's relative growth rate systematically decreases, associated with a shift away from the critical singularity.

Consequently, the von Foerster et al. [29] Doomsday population catastrophe is inherently avoided. The central part of Figure 3 shows a fair portrayal of $P(t)$ data via dependencies included in Eq. (10), using parameters derived and presented in the inset.

### 4. Conclusions

The report presents an analysis of population growth during the Anthropocene, from its onset at 10,000 BCE to 2024. It relies on 'empirical' population data, rather than assuming a heuristic scaling model or using subjectively selected factors. The report presents the implementation of the analytical counterpart [47,51] of the per capita relative growth rate (RGR) concept (Equation 2), first proposed by Lehman et al. [12] in 2021 and further developed in ref. (2025, [13]). It enabled a subtle, distortion-sensitive insight into population data, as shown in Figures 2 and 3. These tests are based on a unique analytical set of 206 population data, presented in Figure 1.

First, the report shows that global population change in the mentioned period is explicitly non-monotonic, as evidenced in Figure 2. Periods related to Malthusian scaling (Eq. 5) are accompanied by 'boosted periods' fairly portrayed by von Foerster et al. [29] (Eq. 1), which can be named a critical-like behavior. In such a case, the system's carrying capacity (Eqs. 8 & 9) increases qualitatively with each human generation, significantly exceeding the needs of subsequent generations. One can expect this process to be essentially innovation-driven, particularly during the Industrial Revolution. In the feedback impact, it promotes the super-Malthusian population growth. Finally, the surplus of the system's Carrying Capacity (Eqs. 8 & 9) for the given epoch can be exhausted, and the period of a lower-level, basic Malthusian growth can appear. Such repeated cycles are visible in Figure 2.

For the feedback coupling between the human population and the carrying capacity developmental patterns, the still ongoing Industrial Revolution (IR) era is exceptional, as evidenced in Figs. 2 and 3. The onset of this era is often linked to the discovery and initial implementations of the steam engine by Thomas Newcomen for removing water from mines in 1712, or with James Watt's discovery (1776) of the improved steam engine concept, which is still in use today [65,66].

However, Figure 2 indicates that the years 1680-1690 may mark the onset of the Industrial Revolution, at least with respect to global population changes. At that time, the Scientific Method for understanding the natural world became widely recognised. In 1687, Isaac Newton published the groundbreaking monograph *Philosophiae Naturalis Principia Mathematica*, which demonstrated the extraordinary power of this new method of cognition [67]. Its thesis remains a significant part of Physics to this day. For the Scientific Method, experimental validation, supported by empirical data analysed via functional scaling relations, is essential. To clarify the fundamental basis of scaling relations, Newton developed the concepts of derivatives and differential equations. However, Newton's intellectual influence started more than two decades earlier [67]. Newton is one of the giants of Science and Philosophy, who shaped the principles of the Scientific Method in the 17$^{th}$ century. They are Galileo Galilei, Johannes Kepler, René Descartes, Francis Bacon, John Locke, and others [67]. Coupling this new, remarkably effective cognitive method with the social transformations and trends of the Enlightenment epoch stirred social structures, leading to a rising wave of innovations, in a feedback coupling with entrepreneurship. The global Carrying Capacity began to grow, as did the population itself.

This extraordinary growth, described in Eq. (1), continued until around 1966, when a qualitative change in the RGR factor occurred. As shown in Figs. 2 and 3 and in Eq. (10), it can be concluded as follows:

$$\sim 1680 \;\rightarrow\; \textcolor{blue}{\textit{critical pattern}} \;\Rightarrow\; \sim 1966 \;\Rightarrow\; \textcolor{red}{\textit{reversed critical pattern}} \quad (11)$$

One of the basic objections to global population growth modelling via the 'Doomsday' Equation (1) was blamed for a 'strange' prediction of the human population extinction when approaching 2026, the year which was deemed accidental and meaningless. Today, one can expect this year to be marked by two history-making events: the end of the War in Ukraine and, potentially, finally, the real end of the War in the Middle East.

Regarding the 'hidden' reversed singularity at $D_2 = 1919 \pm 5$, it can also be considered accidental. However, notable that there is a correlation with the end of World War I. It resulted in the collapse of the major post-feudal Prussian-Germanic, Austrian-Hungarian, and Russian Empires [68]. The world, organizationally, socially, and politically, has definitively begun to seek alternative paths of development [68]. But one could also consider the connection between the 'reversed' singularity year and Edwin Armstrong's discovery of the 'feedback circuit' (1914-1922, [69]), which underlies modern electronics and can also be considered a precursor to Artificial Intelligence (AI). By the second half of the 1960s, this concept had already been implemented on a mass scale. near the crossover indicated in Figures 2 and 3, also using integrated systems that were first presented in 1960 [70]. All the formal conditions for a major revolution leading to new-generation computers, information and communication technologies, the internet, and ultimately the emergence of artificial intelligence were in place.

The 'reversed critical' trend in global population changes, coupled with the evolution of carrying capacity (see Eq. 9), appears to be a consequence of the generic dynamics of the highly complex global system related to 'human Real Intelligence', which emerged during the Industrial Revolution era associated with continuously complexing global networking. It may also be related to global awareness of constraints on energy, pollution, and global warming, as well as global networking, which has found personification in the development of artificial intelligence today. The late 1960s also saw essential cultural changes that shaped the way of life of subsequent generations. It is also the time of a voyage to the Moon, a TV broadcast worldwide. This may also have been a global awareness of Earth's planetary spatial constraints.

This report provides a new cognitive perspective on global population changes, resulting exclusively from empirical data reference, and directly coupled to the system (global) carrying capacity changes. Notably, the uniqueness of the Industrial Revolution epoch is highlighted by the excellent model parameterization via von Foerster et al. [29] 'Doomsday' critical-type Eq. (1) for the first 290 years, and then the crossover to the 'reversed criticality' pattern in the next 6 decades. It can be considered a pattern of human Real Intelligence (RI) evolution, during the epoch of exceptional rise in linkages and complexity, allowing RI to be treated as the model counterpart for AI development.

The implications for both RI's and AI's future are relatively positive. Doomsday singularity can be avoided due to the generic constraints of the system in which RI and AI develop.

## 5. Methods for empirical data analysis

To conduct a local, distortion-sensitive analysis of the RGR factor defined by Eq. (2), it is necessary to obtain an analytical dataset of 'empirical' data on global population changes. This cannot be achieved by simply collecting values from various reference studies and databases, since different estimates of $P(t)$ emerge for the same period, and the further back in time one goes. This is not surprising, given the limited number of references for many regions of the world. This issue even affects places with relatively abundant written sources, such as Europe or China, from at least the 15$^{th}$–16$^{th}$ centuries onwards. In prehistoric times, this was a qualitative problem. To overcome this fundamental challenge in demographic studies of the global population, it has been proposed that global population data be collected from various sources [10,19,71-77], with values indicating a 'gross error' rejected and an 'ad hoc' estimate. This is followed by a set of numerical fittings using the Savitzky–Golay principle [78,79] until an analytical set of $P(t)$ values is obtained — that is, a set of data that can be subjected to the differential analysis necessary to calculate subsequent RGR values (see Eq. (2)). This protocol has been implemented in refs. [13,47,51] for testing the possibility of the Super-Malthusian and Verhulst-type descriptions of the global population. for testing the possibility of Super-Malthusian and Verhulst-type descriptions of the global population. For this report, an improved, unique analytical setup containing 206 global population data points since 10,000 BCE (the onset of the Anthropocene) has been prepared. These values are provided in the Supplementary Information and shown graphically in Figure 1.

**Acknowledgments:** The authors are grateful to IHPP PAS for the possibility of working on the given topic.

**Funding declaration:** The authors are grateful to the National Science Center (NCN, Poland), grant NCN OPUS 2022/45/B/ST5/04005, headed by Sylwester J. Rzoska, for their support.

**Authors contribution:** A. Drozd-Rzoska was responsible for methodology, conceptualization, data analysis, supervision, visualization, and writing; A.A. Sojecka was responsible for conceptualization of socioeconomic and historical issues, related supervision and visualization, and writing. S.J. Rzoska was responsible for funding acquisition, support in data analysis and writing.

**Conflict of Interests:** The authors declare no conflict of interest

**Supplementary Information** is available for this report.

**Data Availability**: Relevant global population data are given in the Supplementary Information.

**Correspondence** and requests for materials should be addressed to the following e-mails:
*A. Drozd-Rzoska: Ola.DrozdRzoska@gmail.com;
*A.A. Sojecka: agata.angelika.sojecka@gmail.com
* S.J. Rzoska: sylwester.rzoska@unipress.waw.pl